\documentclass[twocolumn]{aastex63}
\def\arcsec{$^{\prime\prime}$} 
\newcommand{\msun}{$M_{\sun}$}

\newcommand{\msunyr}{\msun\,yr$^{-1}$}
\newcommand{\halpha}{H$\alpha$}

\newcommand{\mdot}{$\dot{M}$}
\newcommand{\ri}{R$_{\rm i}$}
\newcommand{\rw}{W$_{\rm r}$}
\newcommand{\tmax}{T$_{\rm max}$}

\usepackage{graphicx}
\usepackage[caption=false]{subfig}
\usepackage{xcolor}
\usepackage{url}
\usepackage{hyperref}
\usepackage{comment}

\newcounter{column_number}
\shortauthors{Espaillat et al.}
\shorttitle{Dust Depletion in a Misaligned Disk}
\begin{document}

\title{Evidence for Dust Depletion in a Misaligned Protoplanetary Disk with {\it JWST}}

%% Use \author, \affil, plus the \and command to format author and affiliation 
%% information.  If done correctly the peer review system will be able to
%% automatically put the author and affiliation information from the manuscript
%% and save the corresponding author the trouble of entering it by hand.
%%
%% The \affil should be used to document primary affiliations and the
%% \altaffil should be used for secondary affiliations, titles, or email.

%% Authors with the same affiliation can be grouped in a single
%% \author and \affil call.

\correspondingauthor{Catherine C. Espaillat}
\email{cce@bu.edu}

\author[0000-0001-9227-5949]{C. C. Espaillat}
\affil{Department of Astronomy, Boston University, 725 Commonwealth Avenue, Boston, MA 02215, USA}
\affil{Institute for Astrophysical Research, Boston University, 725 Commonwealth Avenue, Boston, MA 02215, USA}

\author[0000-0003-4507-1710]{T. Thanathibodee}
\affil{Department of Astronomy, Boston University, 725 Commonwealth Avenue, Boston, MA 02215, USA}
\affil{Institute for Astrophysical Research, Boston University, 725 Commonwealth Avenue, Boston, MA 02215, USA}

\author[0000-0003-3616-6822]{Z. Zhu}
\affil{Department of Physics and Astronomy, University of Nevada, Las Vegas, 4505 S. Maryland Pkwy, Las Vegas, NV 89154, USA} 
\affil{Nevada Center for Astrophysics, University of Nevada, 4505 S. Maryland Pkwy., Las Vegas, NV 89154-4002, USA} 

\author[0000-0001-5008-2794]{I. Rabago}
\affil{Department of Physics and Astronomy, University of Nevada, Las Vegas, 4505 S. Maryland Pkwy, Las Vegas, NV 89154, USA} 
\affil{Nevada Center for Astrophysics, University of Nevada, 4505 S. Maryland Pkwy., Las Vegas, NV 89154-4002, USA} 

\author[0000-0002-6808-4066]{J. Wendeborn}
\affil{Department of Astronomy, Boston University, 725 Commonwealth Avenue, Boston, MA 02215, USA}
\affil{Institute for Astrophysical Research, Boston University, 725 Commonwealth Avenue, Boston, MA 02215, USA}

\author[0000-0002-3950-5386]{N. Calvet}
\affil{Department of Astronomy, University of Michigan, 1085 South University Avenue, Ann Arbor, MI 48109, USA} 

\author[0000-0001-9219-7696]{L. Zamudio-Ruvalcaba}
\affil{Department of Astronomy, Boston University, 725 Commonwealth Avenue, Boston, MA 02215, USA}
\affil{Institute for Astrophysical Research, Boston University, 725 Commonwealth Avenue, Boston, MA 02215, USA}

\author[0009-0005-4517-4463]{M. Volz}
\affil{Department of Astronomy, Boston University, 725 Commonwealth Avenue, Boston, MA 02215, USA}
\affil{Institute for Astrophysical Research, Boston University, 725 Commonwealth Avenue, Boston, MA 02215, USA}

\author[0000-0001-9301-6252]{C. Pittman}
\affil{Department of Astronomy, Boston University, 725 Commonwealth Avenue, Boston, MA 02215, USA}
\affil{Institute for Astrophysical Research, Boston University, 725 Commonwealth Avenue, Boston, MA 02215, USA}

\author[0000-0003-1878-327X]{M. McClure} 
\affil{Leiden Observatory, Leiden University, PO Box 9513, NL–2300 RA Leiden, The Netherlands}

\author[0000-0002-3883-9501]{J. F. Babb}
\affil{Center for Astrophysics \textbar\ Harvard \& Smithsonian, Cambridge, MA 02138, USA}

\author[0000-0002-1650-3740]{R. Franco-Hern{\'a}ndez}
\affil{Instituto de Astronom{\'i}a y Meteorolog{\'i}a, Universidad de Guadalajara, Avenida Vallarta No. 2602, Col. Arcos Vallarta Sur, CP 44130, Guadalajara, Jalisco, Mexico}

\author[0000-0003-1283-6262]{E. Mac{\'i}as}
\affiliation{European Southern Observatory, Karl-Schwarzschild-Str. 2, 85748, Garching bei Munchen, Germany}

\author[0000-0003-1621-9392]{M. Reynolds}
\affil{Department of Astronomy, Ohio State University, 140 West 18th Ave., Columbus, OH 43210, USA}
\affil{Department of Astronomy, University of Michigan, 1085 South University Avenue, Ann Arbor, MI 48109, USA} 

\author[0000-0003-1623-1391]{P. -G. Yan}
\affil{Center for Astrophysics \textbar\ Harvard \& Smithsonian, Cambridge, MA 02138, USA}

\begin{abstract} %limit is 250 words 

Here we report the detection of dust depletion in a misaligned inner disk around UX Tau A using {\it JWST} MIRI spectra.  Mid-infrared (MIR) continuum ``seesaw'' variability was detected in this disk by {\it Spitzer} and attributed to variable shadows cast on the outer disk by the inner disk.  The {\it JWST} MIRI spectrum of UX Tau A also shows seesaw variability, but with a significant decrease of emission shortwards of 10 {\micron} to nearly photospheric levels.   We argue that UX Tau A's MIR continuum variability is due to depletion of dust in a misaligned inner disk. We speculate that this dust depletion occurs because the inner disk is misaligned from the outer disk, which can disrupt the replenishment of the inner disk from the outer disk. Using contemporaneous measurements of the mass accretion rate of UX Tau A and estimating the amount of dust necessary to produce the MIR excess in the {\it Spitzer} observations, we estimate a minimum dust depletion timescale of $\sim$0.1 yr. These observations show that we can indirectly detect the signatures of misaligned inner disks through MIR continuum variability and that in some cases the inner disk may be significantly depleted of dust and become optically thin.

\end{abstract}


\section{Introduction} \label{intro}

Variability is a distinctive characteristic of low-mass ($<$ 2 {\msun}), accreting pre-main sequence stars (i.e., Classical T Tauri stars; CTTS). Extensive multi-epoch studies of CTTS in the X-ray, UV, optical, and IR wavelengths revealed significant variability at each of these wavelengths \citep[e.g.,][]{stauffer15}. 

Out of the wide range of variability seen to date, at least two types are linked to dust in the inner regions of the protoplanetary disk: MIR ``seesaw'' continuum variability and optical ``dipper'' light curves. Seesaw variability was discovered by {\it Spitzer}, which found that the MIR continuum in many disks with large central cavities seesaws, i.e., the flux at shorter wavelengths varies inversely with the flux at longer wavelengths \citep{muzerolle09, espaillat11, flaherty11}. Spectral energy distribution (SED) modeling showed that this seesawing behavior could be explained by differences in the size of the shadow cast on the edge of the outer disk by the inner disk. \citet{espaillat11} attributed the seesaw behavior to changes in inner disk wall height resulting in differences in illumination in the outer wall, a geometrical effect. NIR imaging of dark regions supports shadowing of the outer disk by the inner disk \citep[e.g.,][]{facchini18,zhu19,nealon20, benisty23}. ``Dippers'' were identified in {\it Kepler} surveys based on their optical light curves, which display deep dips in their brightness \citep{cody14,stauffer15}.  The period of the dips is typically a few days and equals the stellar rotation period, which indicates that there is dust located at the corotation radius  that obscures the star \citep{bouvier07, cody18}.  

Here, we study UX Tau A, which has been reported to display seesaw variability \citep{espaillat11}, to further explore the role of dust in the innermost disk in MIR continuum variability and the link to dippers and disk shadows.  UX Tau is a quadruple system located in the Taurus-Auriga star-forming region at a distance of $\sim$147 pc \citep{gaia18b}. The UX Tau system is made up of components A, B, and C where A and B are separated by 5.86{\arcsec} and A and C are separated by 2.63{\arcsec} \citep{white01}.  UX Tau B is a close binary separated by 0.1{\arcsec} \citep{duchene24}.  In this work, we focus on the A component of the UX Tau system, which has been shown to dominate the emission \citep{white01}.

UX Tau A has a stellar mass of 1.4 {\msun} \citep{zapata20}, and reported spectral types of G5--K5 \citep{rydgren76, espaillat10, kraus09, hartigan89} with an accretion rate of 1$\times$10$^{-8}$ {\msunyr} \citep{espaillat10}.  This object is surrounded by a pre-transitional disk \citep[i.e., an inner disk separated from an outer disk by a large gap $\sim$\,tens of astronomical units wide;][]{espaillat14}. (Sub-)mm imaging finds that the disk has a gap of $\sim$25--31 au \citep{andrews11,francis20} as well as spiral arm structure indicating that UX Tau A has been perturbed by UX Tau C within the past 1000 yrs \citep{menard20, zapata20}. NIR observations detect dust down to at least 23 au  \citep{tanii12}. 

Here we present new {\it JWST} data of UX Tau A and contemporaneous optical spectra, which we compare to archival MIR spectra and photometry as well as contemporaneous optical light curves.  In Section~2, we present the new and supporting datasets.  In Section~3, we analyze the optical light curves and {\halpha} profiles.  We discuss possible connections between the datasets in Section~4 and end with a summary in Section~5.

%%%%%%%%%%%%%%%%%%%%%%%%%%
%FIGURE 1 - JWST spectrum
\begin{figure*}    
\epsscale{2.4}
\plottwo{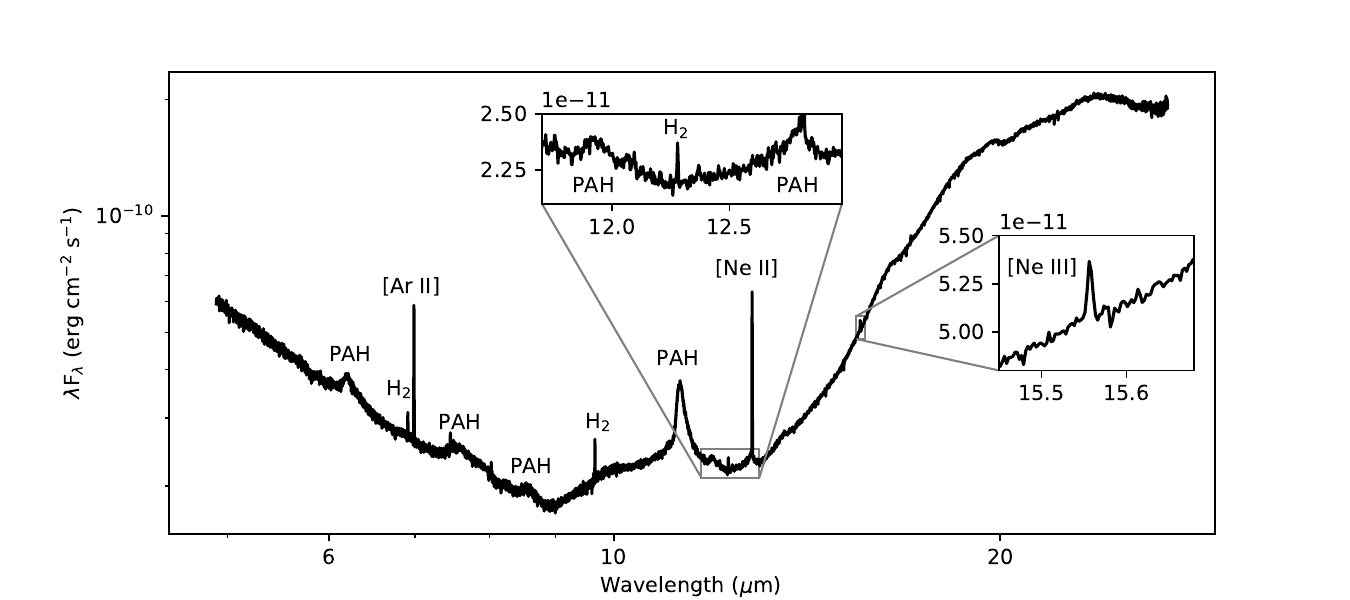}{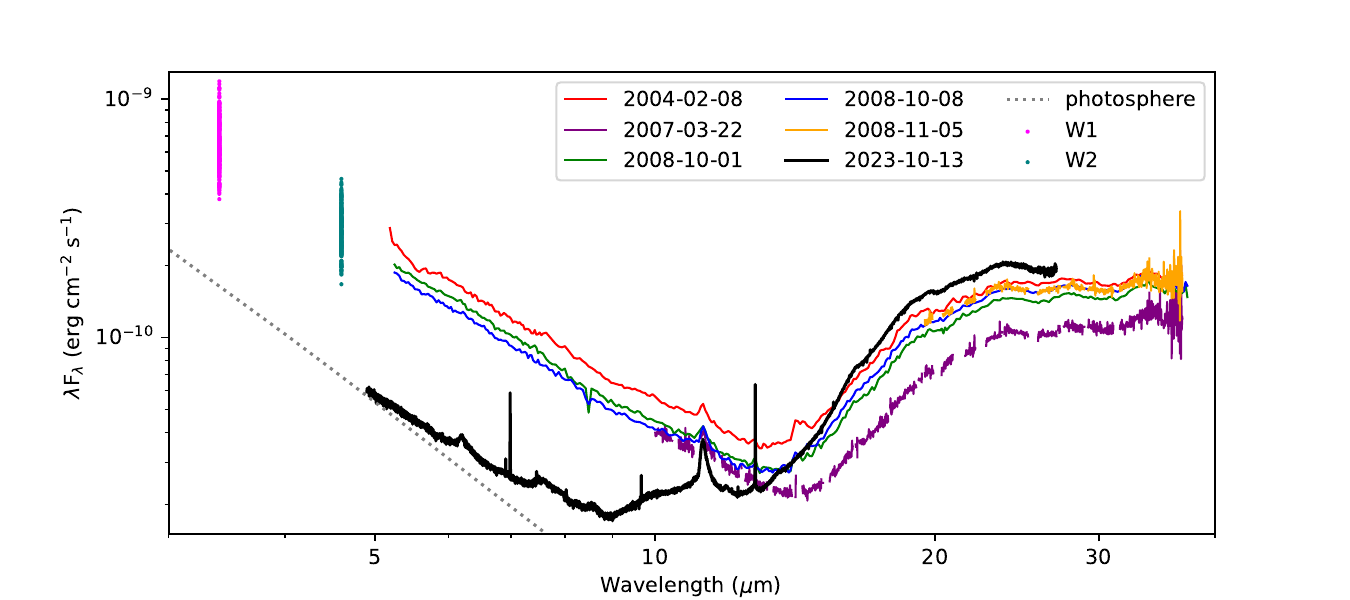}
\caption{
Top: {\it JWST} spectrum of UX Tau A with some atomic lines, H$_{2}$ lines, and PAH features identified. 
Bottom: MIR spectra of UX Tau A displaying MIR continuum variability over $\sim$19 yrs (see legend). The {\it JWST} MIRI MRS spectrum is shown in black, and the other spectra are from {\it Spitzer}. We include WISE/NEOWISE W1 and W2 photometry, which are shown as magenta and teal circles. The stellar photosphere (dotted line) is adapted from \citet{espaillat11}. 
}
\label{fig:JWST}
\label{fig:MIRspectra}
\end{figure*} 
%%%%%%%%%%%%%%%%%%%%%%%%%%

%%%%%%%%%%%%%%%%%%%%%%%%%%
%FIGURE 2 - MIR photometry
\begin{figure*}    
\epsscale{0.9}
\plotone{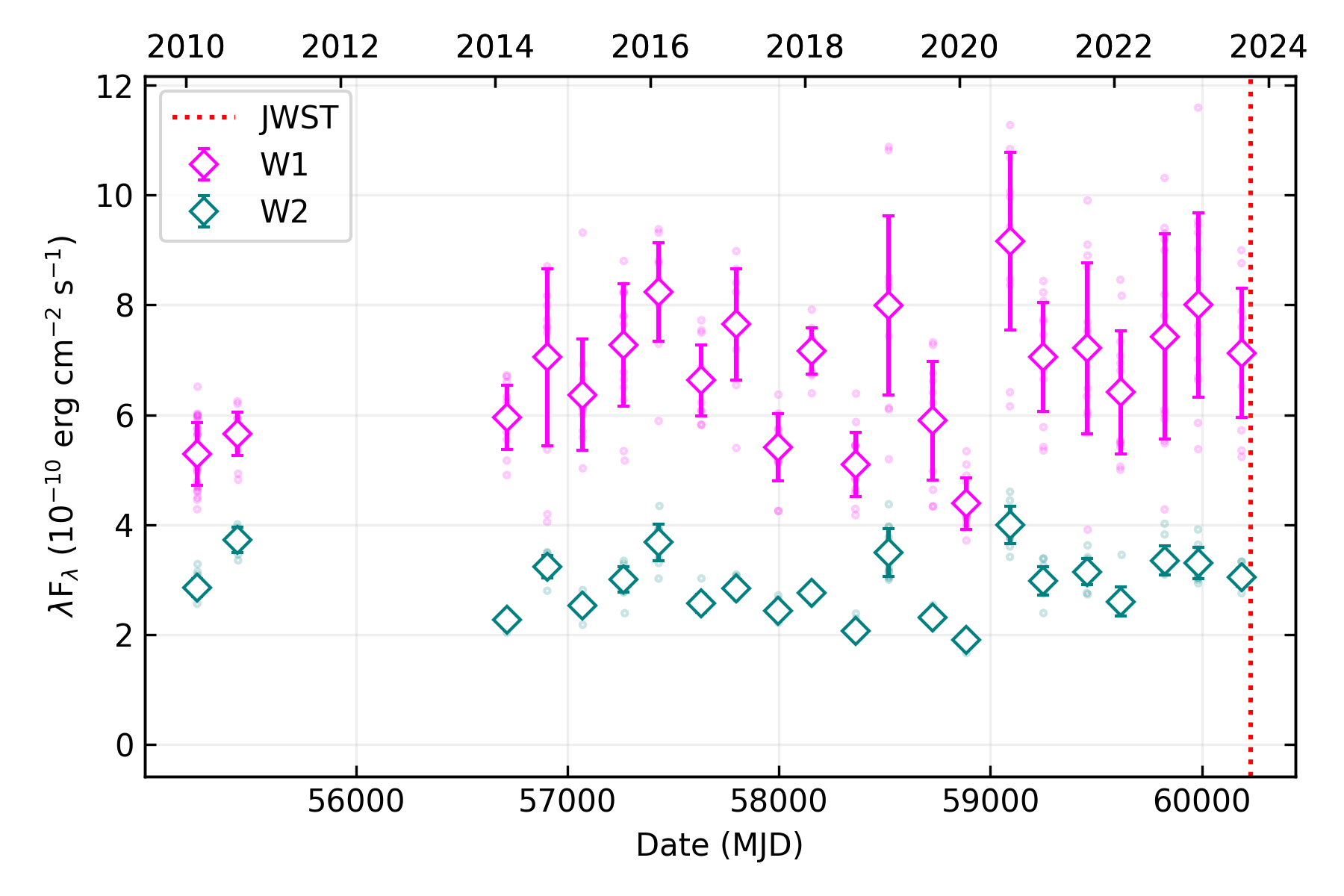}
\caption{
MIR light curve of UX Tau A from WISE/NEOWISE. The W1 and W2 data points are shown as light magenta and teal circles.  The diamonds are the median flux taken over 2--3 days and the error bars are the standard deviation. The vertical broken line notes the time of the {\it JWST} observation.
}
\label{fig:neowise}
\end{figure*} 
%%%%%%%%%%%%%%%%%%%%%%%%%%

\section{Observations \& Data Reduction} \label{redux}

We study MIR spectra from {\it JWST} and {\it Spitzer}, MIR photometry from WISE/NEOWISE, optical light curves from {\it TESS} and ASAS-SN, and optical spectra from LCOGT/NRES.  In the following, we provide more details about these data.

\subsection{Mid-Infrared Data}

\subsubsection{JWST}

We present new {\it JWST} MIRI \citep[][]{reike15, wright23} MRS \citep[][]{wells15} data of UX~Tau A taken on 2023 October 13 starting at 05:37:37 UT as part of GO program 1676 (PI: Espaillat). The observations were 336~s long and used the default four-point dither.  Background observations were also taken with the same exposure time and setup.  We follow the same reduction procedure as in \citet{espaillat23}.  In sum, we reduce the uncalibrated raw MRS data using the calibration reference file version jwst\_1252.pmap and the {\tt JWST} Science Calibration Pipeline v.1.15.1 \citep{bushouse24_v1.15.1}.  There is a known flux calibration uncertainty of $\sim$10$\%$ (Law et al in prep). The spectrum is presented in Figure~\ref{fig:JWST}.  
  
The Channel 1 spectra (4.9--7.5 {\micron}) are of UX Tau A alone while Channels 2, 3 and 4 (7.5--27.90 {\micron}) include the C component. At K-band, the flux ratio between A and C is 16.9$\pm$1.6 and between A and B it is 3.91$\pm$0.31 \citep{white01}. If there was a significant contribution from the C component, there would have been a sharp increase in flux at 7.5 {\micron}. We conclude that the C component is not contributing significantly to the MIR.

UX Tau A has detections of [\ion{Ne}{2}] at 12.81 {\micron}, [\ion{Ne}{3}] at 15.5 {\micron}, and [\ion{Ar}{2}] at 6.98 {\micron}.  There are also a few H$_{2}$ lines present and weak PAH features at $\sim$6.2, 7.7, 8.6, 12.0, and 12.7  {\micron} along with a strong PAH feature at 11.3 {\micron}. 
We follow \citet{espaillat23} to measure [\ion{Ne}{2}], [\ion{Ne}{3}], and [\ion{Ar}{2}] lines fluxes of 
1.5$\pm$0.1$\times$10$^{-14}$,
1.4$\pm$0.3$\times$10$^{-15}$, 
and 1.1$\pm$0.1$\times$10$^{-14}$ erg cm$^{-2}$ s$^{-1}$.
We note that the [Ne II] line flux is consistent with the line flux measured by \citet{szulagyi12}.

\subsubsection{Spitzer} 

In Figure~\ref{fig:MIRspectra}, we show {\it Spitzer} IRS spectra taken from the Combined Atlas of Sources with Spitzer IRS Spectra \citep[CASSIS,][]{lebouteiller15}.  These include low-resolution data of UX Tau A on 2008-10-01 and 2008-10-08. Also included are high-resolution spectra of UX Tau A on 2007-03-22 and 2008-11-05.  We note that the spectrum of UX Tau A from 2004-02-08 appeared to have scaling issues with the CASSIS reduction so here we use the spectrum from \citet{espaillat11}. The systematic absolute flux uncertainties on the IRS spectra are $\sim$5$\%$.

\subsubsection{WISE/NEOWISE} 

In Figure~\ref{fig:neowise}, we use data from from the {\it Wide-field Infrared Survey Explorer} ({\it WISE}) mission \citep{wright10} and from the NEOWISE project \citep{mainzer11}. These data are in {\it WISE} bands W1 (3.4 {\micron}) and W2 (4.6 {\micron}). The WISE data are from mid-February 2010 and mid-August 2010. The NEOWISE data span 2014 February -- 2023 August and are typically obtained in 2--day segments, with 10--20 observations per filter and 130--200 days between segments. The flux calibration uncertainties are less than 5$\%$.

\subsection{Optical Data} 

\subsubsection{TESS and ASAS-SN}

\begin{figure*}    
\epsscale{1.0}
\plotone{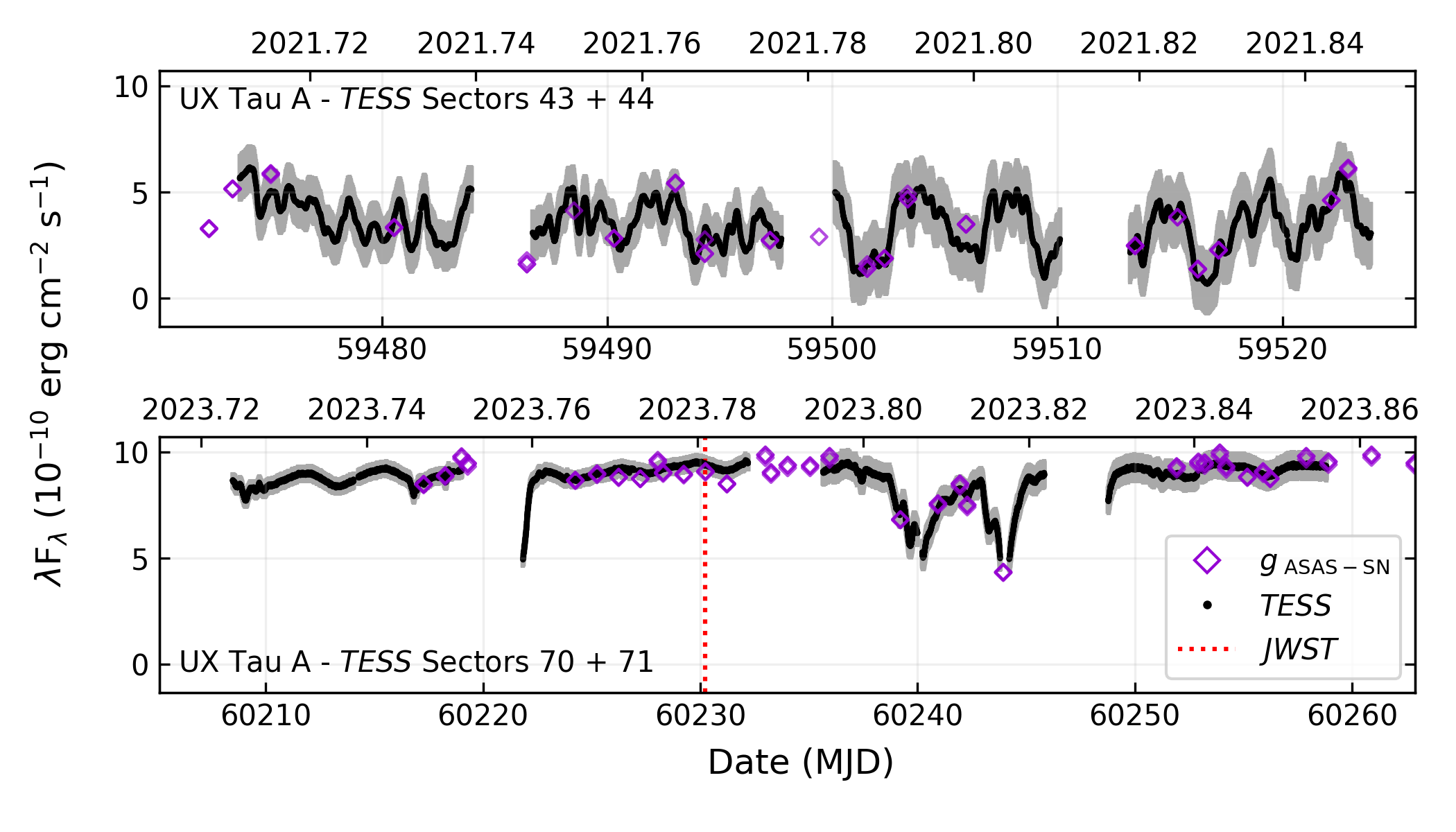}
\caption{
{\it TESS} (black) and ASAS-SN g-band (purple) light curves of UX Tau A.  To facilitate comparison, the {\it TESS} data were scaled to match the ASAS-SN data.  A vertical broken line marks the time of the {\it JWST} observation.  
}
\label{fig:optical}
\end{figure*}

We show All-Sky Automated Survey for Supernovae  \citep[ASAS-SN;][]{kochanek17} and {\it TESS} light curves for UX Tau A in Figure~\ref{fig:optical}. The ASAS-SN $g$ photometry come from Sky Patrol V2.0 \citep{hart23} and are contemporaneous with the {\it JWST} observations. We obtained the {\it TESS} light curves using the {\it TESS-GAIA} Light Curve (tglc) Python package \citep{han23}.

The {\it TESS} observations were simultaneous with {\it JWST}. {\it TESS} observed UX Tau A for $\sim$26 days in Sector 70 from 2023 September 20 to October 16 (MJD: 60207.9-60233.3) with 3.3-minute cadence. In Figure~\ref{fig:optical}, we include additional {\it TESS} light curves of UX Tau A from Sectors 43, 44, and 71 with cadences ranging from 3.3--30 minutes.

%%%%%%%%%%%%%%%%%%%%%%%%%%
%FIGURE 3 - optical photometry
%%%%%%%%%%%%%%%%%%%%%%%%%%

\subsubsection{LCOGT} 

We observed UX Tau A using the NRES spectrograph on the Las Cumbres Global Observatory 1-meter telescope network (LCOGT). NRES is a robotic, fiber-fed spectrograph providing high-resolution (R$\sim$48000) spectra from the BANZAI-NRES automatic reduction pipeline. Multiple exposures were stacked. All spectroscopic observations are contemporaneous with the {\it JWST} observations. UX Tau A was observed starting on 2023-10-11UT00:52, 2023-10-11UT23:50, 2023-10-14UT09:33, and 2023-10-17UT11:20. 
All spectra were re-normalized using a polynomial fit to the continuum and are shown in Figure~\ref{fig:halpha}.

\section{Analysis \& Results} 

We measure the periods of the optical light curves and classify them following \citet{cody14}.  Then we model the {\halpha} profiles to measure accretion rates and properties of the accretion flow. These results will be discussed along with the MIR data in Section~4.

\subsection{Optical light curves}

To measure the periods in the {\it TESS} data, we subtract a linear fit from each light curve and fit a Lomb-Scargle periodogram. We measure the $Q$ and $M$ variability metrics following \citet{cody18}. $Q$ measures the light curve's periodicity and varies between 0 and 1 where 0 is periodic and 1 is aperiodic. Q values between 0.15 and 0.85 are quasiperiodic. $M$ measures the light curve's asymmetry and typically varies between -1 and 1.  Positive values correspond to dips and negative values to bursts with the highest values corresponding to dippers and the lowest values to bursters.   M values between $\pm0.25$ are symmetric.

The period of UX Tau A is $\sim$3.8 days amongst the {\it TESS} sectors (43, 44, 70, and 71). We find Q values of 0.9, 0.4, 0.9, 0.9 and M values of -0.3, 0.2, 0.5, 1.2.  These metrics lead to the following classifications: Sector 43 burster (B); Sector 44 quasiperiodic symmetric (QPS); Sector 70 aperiodic dipper (APD); Sector 71 aperiodic dipper (APD).

%%%%%%%%%%%%%%%%%%%%%%%%%%
%FIGURE 4 - Halpha modeling 
\begin{figure*}    
\epsscale{1.0}
\plotone{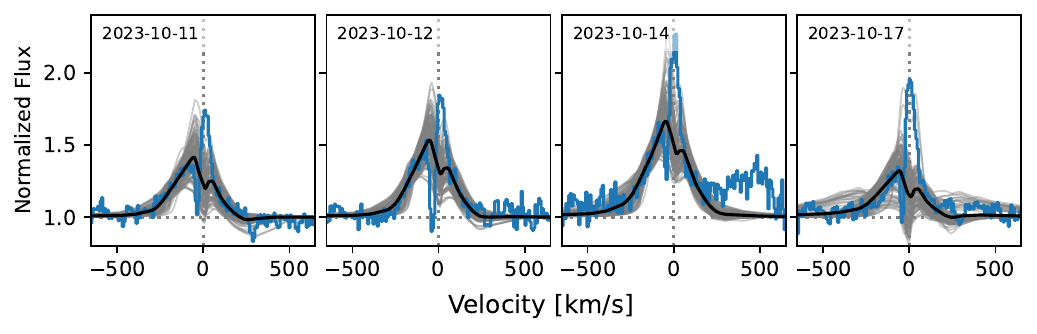}
\caption{
{\halpha} profiles (blue) of UX~Tau~A and the average accretion flow model fit (black line) along with the top 100 best-fitting models (gray lines). The horizontal and vertical dotted lines correspond to the continuum and line center, respectively. 
}
\label{fig:halpha}
\end{figure*} 
%%%%%%%%%%%%%%%%%%%%%%%%%%

%%%%%%%%%%%%%%%%%%%%%%%%%%
%TABLE 1 
\begin{deluxetable*}{cccccccc}
\tablecaption{Results of the Magnetospheric Accretion Model \label{tab:model_results_acc}}
\tablehead{
\colhead{Object} &
\colhead{Obs. Date} &
\colhead{Instrument} &
\colhead{\mdot} &
\colhead{\ri} &
\colhead{\rw} &
\colhead{\tmax} &
\colhead{$i$} \\
\colhead{} &
\colhead{(UT)} &
\colhead{} &
\colhead{($10^{-9}$\,\msunyr)} &
\colhead{(R$_{\star}$)} &
\colhead{(R$_{\star}$)} &
\colhead{($10^3\,$K)} &
\colhead{(deg)}
}
\startdata
UX~Tau~A & 2023-10-11 & NRES  & 16.0$\pm$29.2 & 2.3$\pm$1.6 & 0.4$\pm$0.3 &  9.0$\pm$0.7 & 46$\pm$14 \\
UX~Tau~A & 2023-10-12 & NRES  & 11.7$\pm$22.0 & 2.0$\pm$1.2 & 0.4$\pm$0.3 &  9.0$\pm$0.6 & 46$\pm$13 \\
UX~Tau~A & 2023-10-14 & NRES  & 21.6$\pm$31.7 & 1.7$\pm$0.4 & 0.3$\pm$0.2 &  9.0$\pm$0.6 & 48$\pm$12 \\
UX~Tau~A & 2023-10-17 & NRES  & 11.3$\pm$26.9 & 3.5$\pm$2.2 & 0.7$\pm$0.6 &  8.7$\pm$0.7 & 36$\pm$16 \\
\enddata
\end{deluxetable*}
%%%%%%%%%%%%%%%%%%%%%%%%%%

\subsection{Optical spectra}

We model the {\halpha} profiles of UX Tau A (Figure~\ref{fig:halpha}) using the magnetospheric accretion flow model from \citet{hartmann94,muzerolle98,muzerolle01}. The magnetic, stellar, and disk rotation axes are aligned and the flow geometry follows a dipolar magnetic field which has the following parameters: inner radius (\ri), width of the flow (\rw) at the disk plane, maximum temperature in the flow (\tmax), and viewing inclination ($i$). To determine the best fits, we follow the procedure of \citet{thanathibodee23}. 

We created a grid of 29,700 models with the following values: \mdot~($1\times10^{-9}-1\times10^{-7}$\,\msunyr), \ri~($1.5-7.0$\,R$_{*}$), \rw~($0.2-1.8$\,R$_{*}$), \tmax~($8-10\times10^3$\,K), and $i$~($20\degr-60\degr$). We calculate the $\chi^2$ for each combination of the model and observed profile and we selected the models where the normalized likelihood is $\geq0.5$ and calculated the weighted mean of \mdot, \ri, \rw, \tmax, and $i$.  The best fit parameters are listed in Table~1.

\section{Discussion} 

Seesaw variability is still present in the {\it JWST} spectrum of UX Tau A, but differs significantly from what has been seen previously. Strikingly, the {\it JWST} spectrum is nearly consistent with photospheric emission at the shortest wavelengths, with a very small MIR excess shortwards of 10 {\micron} and a broad, weak 10 {\micron} silicate emission feature.\footnote{The photosphere plotted in Figure~\ref{fig:MIRspectra} uses colors from \citet{kh95}, is scaled at J-band, and follows a Rayleigh-Jeans tail beyond K-band. The photometry used in \citet{espaillat11} included emission from the A, B, and C components and here we scale the photosphere following the flux ratios from \citet{white01} to represent the emission from only the A component.}  This combination points to a small amount of $\sim$micron-sized dust grains in an optically thin inner disk.  Previous work has reproduced the 10 {\micron} feature of UX Tau A using a grain size distribution of a$^{-3.5}$, where $a$ varies between a$_{min}$=0.005 {\micron} and a$_{max}$=10 {\micron} \citep{espaillat11}.

We propose that the drop to nearly photospheric levels in UX Tau A is due to dust depletion in a misaligned inner disk undergoing disk breaking. Disk breaking is a phenomenon that occurs when the inner and outer disks precess independently, possibly due to the presence of a companion \citep{zhu19}. The inclination of the outer disk of UX Tau A is 37.96$^{\circ}$$^{+0.97}_{-0.90}$ while the inclination of the inner disk is 73.46$^{\circ}$$^{+11.76}_{-16.47}$ \citep{bohn22}.  

\begin{figure*}    
\epsscale{1.0}
\plottwo{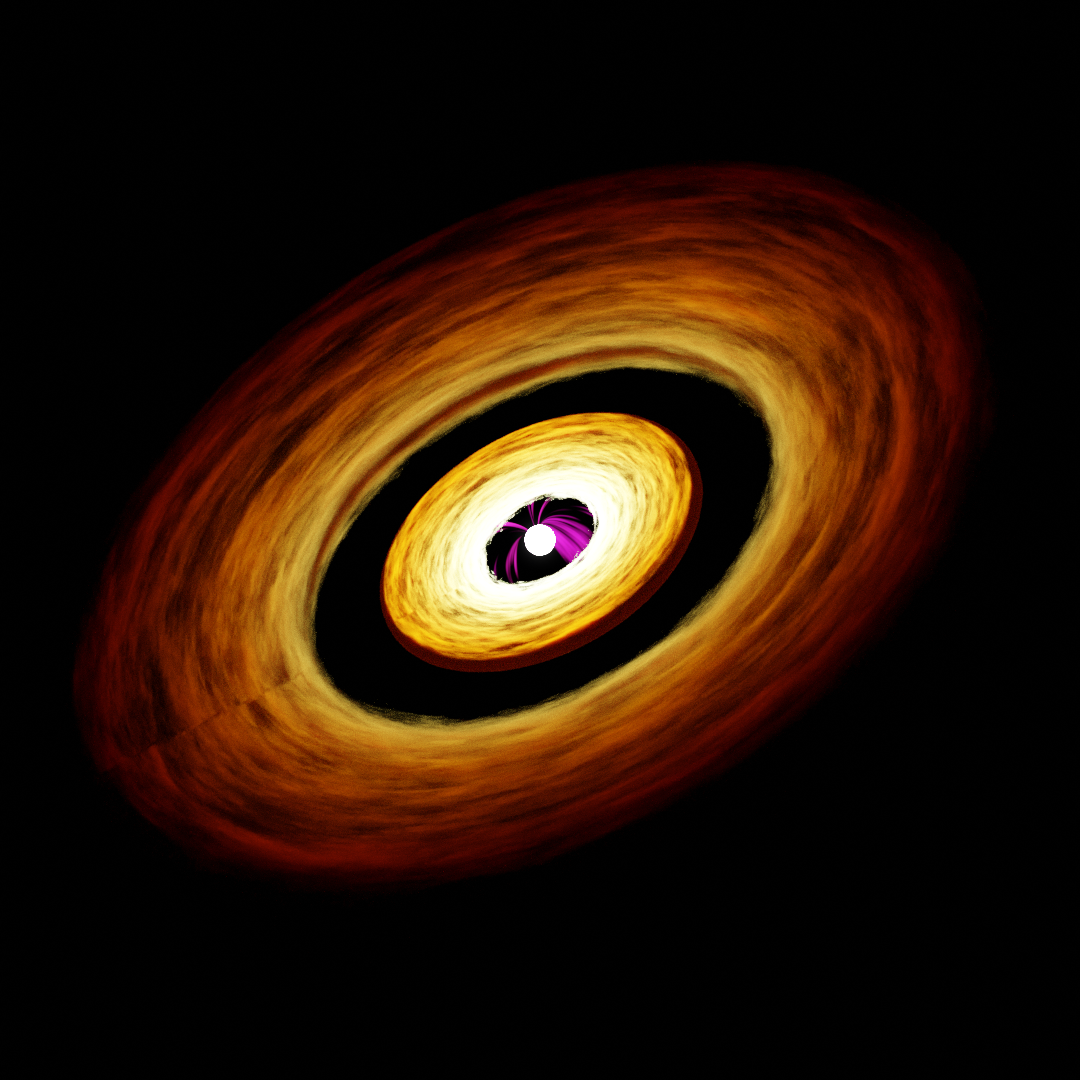}{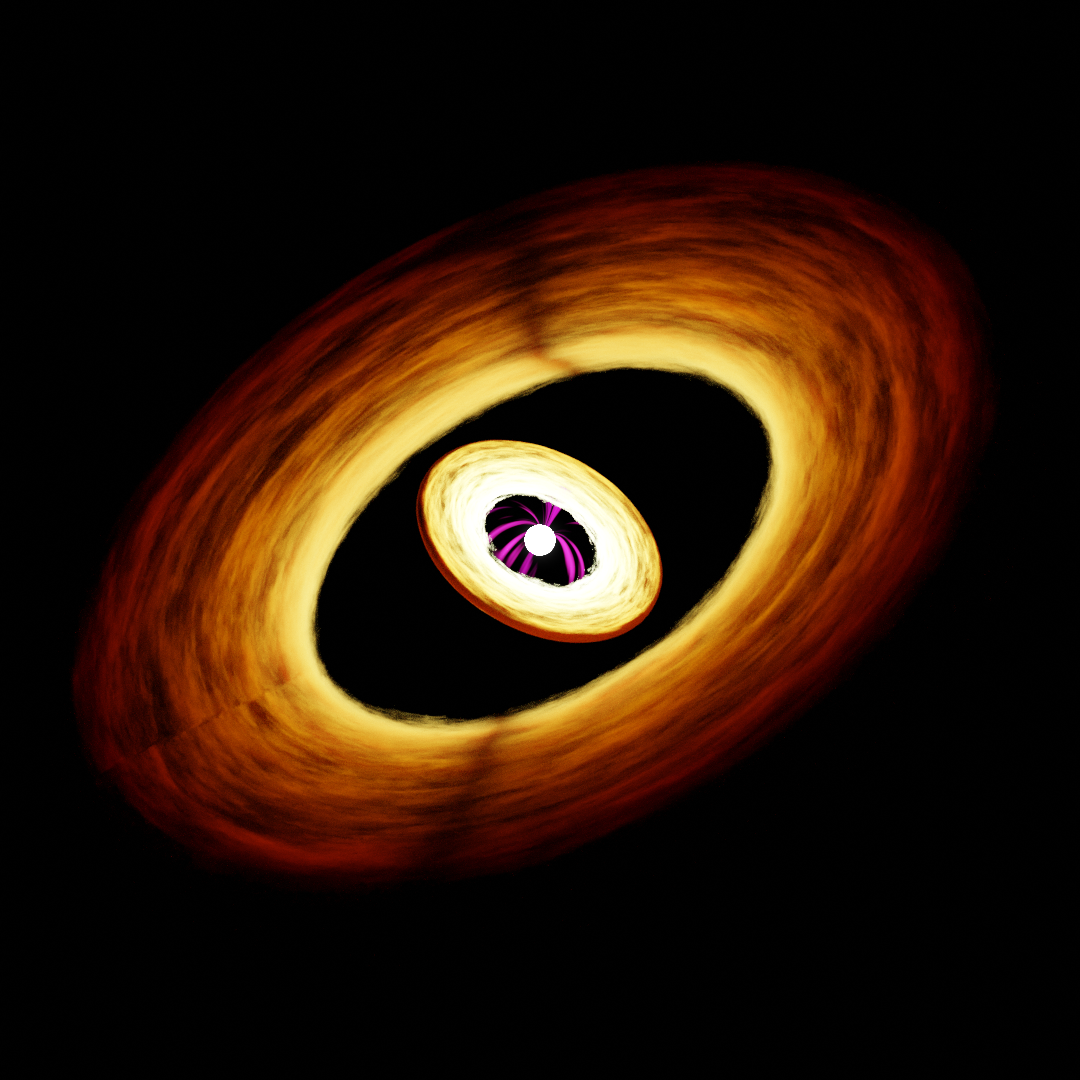}
\caption{
Schematics showing that the inner and outer disks are coplanar (left) and misaligned (right). If the inner disk has nodal precession along the vertical direction, it changes from the left to right panel after half the precession cycle. 
}
\label{fig:schematic}
\end{figure*}

As the inner disk precesses, there will be times when it is aligned with the outer disk (Figure~\ref{fig:schematic}, left) and times when it is misaligned (Figure~\ref{fig:schematic}, right).  Mass transfer from the outer disk to the inner disk is easiest when the disks are coplanar and is the most disrupted when the inner disk is at maximum misalignment from the outer disk.  However, accretion may still occur at intermediate points. Because it only takes a small amount of dust to make the disk optically thick, any accretion can deposit sufficient dust to lead to an optically thick inner disk. Therefore, we expect that a significant MIR excess would be seen most of the time.  Then when the accretion to the inner disk is significantly disrupted, most of the dust in the inner disk may accrete or drift to the star due to lack of replenishment from the outer disk. The inner disk becomes optically thin and there will be a drop at the shorter wavelengths in the {\it JWST} spectrum, while there is an increase at the longer wavelengths as there is little/no shadow cast on the outer disk. Below we estimate the dust depletion and precession timescales and consider this proposed scenario in light of the MIR light curve, optical light curves, and the MIR emission lines and features.

%%%%%%%%%%%%%%%%%%%%%%%%%%
%FIGURE 5 - schematic
%%%%%%%%%%%%%%%%%%%%%%%%%%

\subsection{Depletion and precession timescales}

We can estimate the dust depletion timescale by calculating how much mass was in the inner disk at the time of the {\it Spitzer} observations and comparing this to the accretion rate of UX Tau A. Assuming the excess MIR emission comes from a marginally optically thick inner disk at 1550~K, the lower limit of the dust mass in the inner disk is then $\sim$10$^{-11}$ M$_{\odot}$ using a dust opacity of 200 cm$^2$ g$^{-1}$ \citep{birnstiel18}. We measure accretion rates of 1--2$\times$10$^{-8}$ {\msunyr} contemporaneous with the {\it JWST} observations (Table~1), which would quickly deplete the inner disk of dust in 0.1 yrs.  Here we adopt the typically assumed dust-to-gas ratio of 0.01, but the dust-to-gas ratio in circumstellar disks is unknown and it is not clear what it would be in the inner disk of UX Tau A. 

We can also consider the precession timescale of the inner disk of UX Tau A. Theoretical works show that a misaligned Jupiter-mass planet can drive misalignment and precession of the inner disk at $\sim$1000 times the planetary orbital timescale \citep[e.g.,][]{zhu19}. We assume that the period measured from the {\it TESS} light curves (Figure~\ref{fig:optical}) is the stellar rotation period. Then, if a misaligned Jupiter-mass planet is around the corotation radius, the inner disk's precession timescale is $\sim$500 wks or 10 yrs. If the planet is 10 Jupiter masses, the precession timescale is then 1 year. 

\subsection{MIR light curve}

There are no significant dips in the WISE/NEOWISE fluxes.  Our estimated dust depletion timescale is $\sim$0.1 yrs ($\sim$37 days) and the dust replenishment timescale would be similar.  Since there are at least 130 days in between the segments of the WISE/NEOWISE data, it is possible that we missed this short-lived depleted period.  If the precession timescale is 10 yrs, the last time that the inner disk was in the same position as  in the {\it JWST} observation would have been in 2013, which falls in the $\sim$3.5 yrs gap in coverage between WISE and NEOWISE.  It is plausible that given the sparse time cadence of the data and the short depletion and replenishment timescale, that another significant MIR dip was not observed. 

There is significant MIR emission in the last NEOWISE observation taken $\sim$43 days before the {\it JWST} observations, which implies a quick drop to nearly photospheric levels. This is still consistent with our estimated depletion timescale. During the last NEOWISE observations, the inner disk could have been marginally optically thick and then it became optically thin when observed with {\it JWST}. However, the estimated depletion timescale is a lower limit and any slightly longer timescale would not be consistent with the observations.

\subsection{Optical light curves}

One may speculate that the drop in the {\it JWST} spectrum is due to an edge-on inner disk. However, we can exclude the possibility that the inner disk of UX Tau A was close to edge-on at the time of the {\it JWST} observations since there is no evidence of extinction of the star by the inner disk in the simultaneous optical light curve (Figure~\ref{fig:optical}). 

The dips in the {\it TESS} light curve suggest that there is dust in the inner disk along our line of sight which obscures the star a few days before and after the {\it JWST} observations.  This is consistent with the {\it JWST} spectrum whose small MIR excess shows that there is still some small amount of dust grains in the inner disk dust.  The exact mechanism by which this dust obscures the star in the UX Tau A system is unclear. Scenarios that have been proposed to explain dippers include dusty accretion flows \citep{nagel24}, disk warps \citep{mcginnis15}, and dusty disk winds \citep{gaidos24}. 

Finally, we see a change in UX Tau A's optical light curve behavior that could possibly be further evidence of a precessing misaligned inner disk.  In 2021 (top, Figure~\ref{fig:optical}), UX Tau A had a quasiperiodic symmetric light curve and two years later (bottom, Figure~\ref{fig:optical}) showed a dipper light curve. This could indicate that the inner disk was at different inclinations in 2021 and 2023, changing the obscuration of the star and leading to the different light curve behavior.  This may have implications on a subset of objects with ``hybrid'' optical light curves which switch from quasiperiodic to dipper \citep{mcginnis15,cody18}.  

\subsection{MIR atomic gas emission lines and PAH features}

The proposed scenario of a depleted inner disk also appears roughly consistent with the lines and features seen in the {\it JWST} spectrum. First, only the 11.3 {\micron} PAH feature was clearly seen in the {\it Spitzer} spectrum \citep{watson09} whereas they are all prominent in the {\it JWST} spectrum. The PAH features were likely ``drowned out'' in the {\it Spitzer} spectrum due to the significant MIR excess. In the {\it JWST} spectrum, the depletion of dust from the inner disk leads to a decrease in the continuum emission, allowing the PAHs to be conspicuous. Likewise, it is easier to see the [Ar II] and [Ne II] emission lines due to the decreased MIR continuum, as has been seen in other pre-transitional and transitional disks \citep{espaillat07a, espaillat13}.  In addition, the presence of the [Ar II] and [Ne II] lines confirms that there is gas in the inner disk, consistent with the measured accretion rates in this work, which points to a significant gas reservoir in the inner disk despite the depletion of dust.

\section{Summary \& Conclusions} 

We presented a new {\it JWST} spectrum of UX Tau A which shows a striking decrease of continuum emission shortwards of 10 {\micron}, dropping to nearly photospheric levels. We conclude that the dust in the inner disk has been depleted to the point where the inner disk is optically thin, leading to this drop in MIR emission. This dust depletion could be due to a disruption in the replenishment of the inner disk since accretion from the outer disk may be more difficult when the inner and outer disks are misaligned.

Monitoring of the precession timescale via inclination measurements, MIR spectra, and/or imaging is necessary to test if this variability is periodic, which would be expected if it is due to disk precession. More detailed modeling with non-axisymmetric disks is also necessary. These approaches will help us link the variability seen in protoplanetary disks at different wavelengths.

\begin{acknowledgments}
We acknowledge support from {\it JWST} grant GO-01676. We thank the anonymous referee for a constructive report which improved the paper.

The {\it JWST} data presented in this paper were obtained from the Mikulski Archive for Space Telescopes (MAST) at the Space Telescope Science Institute. The specific observations analyzed can be accessed via \dataset[10.17909/sej8-g173]{https://archive.stsci.edu/doi/resolve/resolve.html?doi=10.17909/sej8-g173}.
\end{acknowledgments}

\facilities{ASAS-SN, JWST, LCOGT, WISE, NEOWISE, TESS}

%https://aastex.aas.org/ApJL/countwords.html

%\bibliography{bib_2024_06}
%\end{document}

\end{document}